\documentclass[journal]{IEEEtran}
\usepackage{siunitx}
\ifCLASSINFOpdf
\else
\fi
\usepackage{array}
\usepackage{balance}
\usepackage{graphicx}
\usepackage{caption}
\usepackage{subcaption}
\usepackage{wrapfig}
\usepackage{amsmath}
\usepackage{epstopdf}
\usepackage{amssymb}
\usepackage{multirow}
\usepackage{cite}
\usepackage{multicol}
\usepackage{color}
\usepackage[table]{xcolor}
\usepackage{comment}

\usepackage{amssymb}

\usepackage{longtable}
\usepackage[font=footnotesize,labelfont=bf]{caption}
\newcolumntype{P}[1]{>{\centering\arraybackslash}p{#1}}
\usepackage{url}
\usepackage{hyperref}

\usepackage{array}

\usepackage{url}
\usepackage{siunitx}
\usepackage{enumerate}
\usepackage[normalem]{ulem}
\usepackage{epstopdf}

\usepackage{amssymb}
\usepackage{pifont}
\usepackage{cancel}

\newcommand\figref{Figure~\ref}
\begin{document}
	\title{Performance Investigation of p-FETs Based on Highly Air-stable Mono-layer Pentagonal~PdSe$_2$}
	
	\markboth{}%
	{}
	\author{Keshari~Nandan,~\IEEEmembership{Graduate Student Member,~IEEE,} Amit Agarwal, Somnath Bhowmick, and Yogesh~S.~Chauhan,~\IEEEmembership{Fellow,~IEEE}
		\thanks{This work was partially supported by the Humboldt Foundation and the Swarnajayanti Fellowship (Grant No. – DST/SJF/ETA-02/2017-18) of the Department of Science and Technology~(DST), Government of India.}
		\thanks{K.~Nandan and Y.~S.~Chauhan are with the NanoLab, Department of Electrical Engineering, Indian Institute of Technology Kanpur,
			India, 208016, e-mail: keshari@iitk.ac.in, chauhan@iitk.ac.in.}
		\thanks{A.~Agarwal is with the Department of Physics, Indian Institute of Technology Kanpur, India.}
		\thanks{S.~Bhowmick is with the Department of Material Science and Engineering, Indian Institute of Technology Kanpur, India.}
	}
	\maketitle

\begin{abstract}
	Pentagonal~PdSe$_2$ is a promising candidate for layered electronic devices, owing to its high air-stability and anisotropic transport properties. Here, we investigate the performance of p-type FET based on PdSe$_2$ mono-layer using multi-scale simulation framework combining Density functional theory and quantum transport. We find that mono-layer PdSe$_2$ devices show excellent switching characteristics ($<$ 65 mV/decade) for the source-drain direction aligned along both [010] and [100] directions. Both directions also show good on-state current and large transconductance, though these are larger along the [010] direction for a 15 nm channel device. The channel length scaling study of these p-FETs indicates that channel length can be easily scaled down to 7 nm without any significance compromise in the performance. Going below 7 nm, we find that there is a severe degradation in the sub-threshold swing for 4 nm channel length. However, this degradation can be minimized by introducing an underlap structure. The length of underlap is determined by the trade-off between on-state current and the switching performance.
		
\end{abstract}


\section{Introduction}

\IEEEPARstart{T}{he} demonstration of mono-layer molybdenum disulfide ($\rm MoS_2$) based FET by Radisavljevic $\it{et~al.}$ \cite{MoS2_FET} in 2011 opened up the field for exploration of layered materials' based FETs for future technology nodes. Since then, there has been remarkable progress on the synthesis of layered materials and devices based on them. Both the material and device communities are actively exploring novel 2D materials\cite{kostya} and device concepts based on them. Several promising devices viz. two-dimensional electrostrictive field-effect transistor (2D-EFET)\cite{E-FET}, Dirac-source FET (DS-FET) \cite{DS-FET}, and dual transport FET (DT-FET)\cite{DT-FET} have been demonstrated employing these materials properties, in the last five years.

Beyond $\rm MoS_2$ \cite{Sayeef,Andras,Zeng2017}, several other layered materials have been explored for device applications. Most of them are isotropic, but some of them show interesting anisotropic behavior. The isotropic/anisotropic property is a consequence of high/low lattice symmetry.
The anisotropic materials, with hexagonal buckled or puckered structure such as phosphorene\cite{P1,P2,P3,6905726}, silicene\cite{sili1,sili2}, arsenene\cite{ars1,ars2}, germanene\cite{ger2,ger1}, and stanene\cite{sta1}, have great potential for device application, but their realization in a stable form is still a great challenge. Stability is one of the most important parameters for a material to be considered for electronic applications and large-scale production.

Motivated by this, we study the electronic properties and device characteristics of mono-layer PdSe$_2$ based devices. It is another low symmetric material that has no soft modes in their mono- and few- layers' phonon spectrum\cite{oyedele,lan}. Experimentally, it has been shown that the air-stability of pentagonal-PdSe$_2$ is higher ($>$ 60 days) than other puckered and buckled layered materials\cite{oyedele}. The experimental band gap value is $\SI{1.3}{\electronvolt}$ for this mono-layer. This
	band gap value indicates an excellent semiconductor that can absorb light for electronic and
	optoelectronic devices. This value is not far from silicon, and it is satisfactory for logic
	switches to sharply discriminate between two logic states. Till now, only few-layer pentagonal-PdSe$_2$ based FETs have been demonstrated\cite{oyedele,GIUBILEO202050}, and, hence the performance for mono-layer is unknown. Mono-layer of pentagonal PdSe$_2$ has better electrical transport properties for holes than electrons\cite{lan}. Hence, We have chosen to study p-type FET.

In contrast to conventional materials (bulk Si/Ge, III-V), for new materials, the device simulation becomes more challenging as their characterization parameters are not known. Thus, we start from first-principles calculations of the electronic properties and use these to construct tight-binding (TB) Hamiltonian within maximally-localised Wannier functions (MLWFs) basis. Our device simulations are based on the effective TB model\cite{datta2005quantum,luisier,guo,NanoTCAD,priyank,tapas}, which makes them computationally more efficient, compared to first-principles based device simulation approaches. Our device simulations reveal the following:

\begin{enumerate}
	\item  The best performance of our devices meets the high-performance (HP) expections of International Technology Roadmap for Semiconductors (ITRS) 2013 for the year 2028.
	\item  The best performance of our device for low power (LP) fulfils $\sim$ 51\% requirement of ON current mentioned in ITRS 2013 roadmap for the year 2028.
	\item  Our device's best performance is better than Si FinFET (see Table \ref{table4}).
\end{enumerate}

This paper is organized as follows: The methodology is described in section II. Results and discussion on electronic structures and device performance are presented in section III, and the conclusive remarks are given in section IV.

\section{Methodology}
\begin{figure}[!t]
	\centering
	\includegraphics[width=0.5\textwidth]{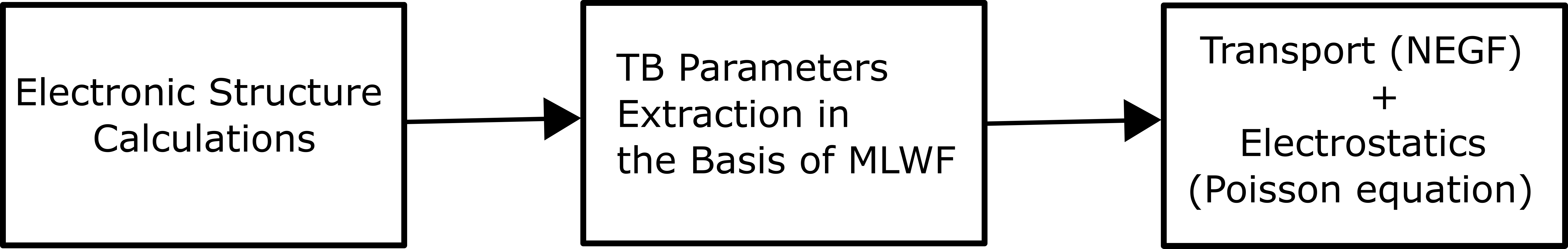}
	\caption{\textbf{Methodology:} Schematic of simulation flow (from DFT to QT).
		It starts with solving the Kohn-Sham equation for a relaxed and periodic system, followed by Wannier transformation and device simulation.}
	
	\label{fig:2}
\end{figure}

\figref{fig:2} shows the simulation methodology used in this work and this methodology is described as follows:
\subsection{Electronic Structure Calculations}
We use Quantum Espresso~(QE)~\cite{QE1,QE2}, a Density-functional theory (DFT) based tool, for electronic structure calculations from first-principles. The Projector Augmented Wave method (PAW) \cite{kjpaw,PAW} with a plane wave basis set is used for the DFT calculations. The exchange correlation effect are approximated by the generalized gradient approximation (GGA) within the Perdew-Burke-Ernzerhof (PBE) formulation \cite{pbe}. The energy cutoffs for wave-functions and charge-density are set to $\rm 60~Ry$ and $\rm 600~Ry$, respectively. The Brillouin zone is sampled using $10\times10\times 1$ Monkhorst-Pack grids \cite{MPgrids}. A $\rm 30~\AA$ thick vacuum layer is used in simulations to avoid the interaction between spurious neighboring layers in the out of plane direction. The primitive unit cell structure is optimized until all components of forces are less than $\rm 10^{-3}$ Ry/a.u.. 

\begin{figure*}
	\centering
	\includegraphics[width=1.0\textwidth]{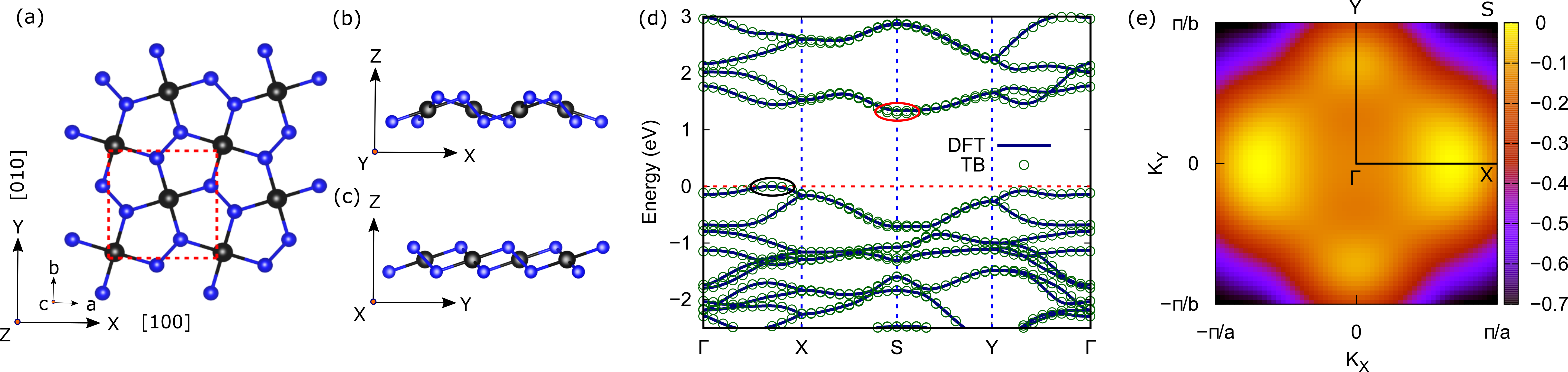}
	\caption{\textbf{Geometric and Electronic Structures :} (a)~Top view, (b)~side view in x-z plane, and (c) side view in y-z plane of pentagonal PdSe$_2$. The Crystallographic directions [100] and [010] are equivalent to X and Y directions of Cartesian coordinate system.
		The black and blue balls represent Pd and Se atoms, respectively. The dotted red color rectangle in (a) is primitive unit cell of the system.
		(d)~Comparison of energy bands calculated from DFT and tight binding Hamiltonian (in MLWF basis) along high symmetry paths ($\rm \Gamma$-$\rm X$-$\rm S$-$\rm Y$-$\rm \Gamma$) in Brillouin zone for mono-layer of pentagonal-PdSe$_2$. VBM is shifted to zero energy. (e) Contour plot of the highest valance band. High symmetry points and transport directions are also indicated in (e).}
	
	\label{fig:1}
\end{figure*}

\subsection{Maximally-localised Wannier Functions (MLWFs) Tight-Binding Hamiltonian}
First-principles calculations provide the electronic structure in terms of Bloch states. The Bloch states corresponding to conduction and valance bands, which lie near the band gap, are mapped to MLWFs using the Wannier90 suite of codes\cite{W90} in continuous space of unitary matrices\cite{W90_theory,MLWF1,MLWF2}. Five d-orbitals and three p-orbitals of each Pd and each Se atom are used to generate an initial guess for mapping of Bloch states, to capture eighteen highest valance bands and four lowest conduction bands. Once the Wannier functions (WFs) are known, the on-site energy and hopping parameters are calculated to generate the elements of Hamiltonian matrix ($ H_{ijk}^{mn}$). The $ H_{ijk}^{mn}$ couples $\rm m^{th}$ Wannier function in the home unit cell (0, 0, 0) to the $\rm n^{th}$ Wannier function in the unit cell (i, j, k). Using unit cell coupling terms, we construct the device Hamiltonian, which is given as input to the NEGF solver.

\subsection{Device Simulations}
At each bias point ($V_{GS}$, $V_{DS}$), Non-equilibrium Green's functions (NEGF) and Poisson equation are solved self-consistently to obtain the converged values of charge, potential, and transmission coefficient for the device. In this formalism, transmission coefficient can be expressed as,

\begin{align}\label{Transmission}
\nonumber
T(E,k,V_{GS},V_{DS}) = Trace[\Gamma_L(E,k,V_{GS},V_{DS}) G^R(E,k,\\V_{GS},V_{DS}) 
\Gamma_R(E,k,V_{GS},V_{DS}) G^A(E,k,V_{GS},V_{DS})]~.
\end{align}
Here, $G^R$ and $G^A = [G^R]^\dagger$ are retarded and advance Green functions, respectively. $\Gamma_{L/R}( = i[\Sigma_{L/R}-\Sigma^\dagger_{L/R}]$) is the broadening from left/right contacts. $\Sigma_{L/R}$ is the left/right contact self-energy matrix. k is the wave-vector in transverse (channel width) direction.

The current is calculated using the Landauer-$\rm B\ddot{u}ttiker$ approach\cite{Landauer}.
The source to darin current ($I_{SD}$) for a given bias ($V_{GS}$, $V_{DS}$) can be expressed as,

\begin{align}\label{Landauer_formulation}
\nonumber
I_{SD}(V_{GS},V_{DS}) = \frac{2q}{h} \int_{-\infty}^{\infty} \sum_{k} T(E,k,V_{GS},V_{DS})\\ [f(E-\mu_D)-f(E-\mu_S)] dE~,
\end{align}
Equation. 2 can also be written as,
\begin{align}
\nonumber
I_{SD}(V_{GS},V_{DS})=\int_{-\infty}^{\infty} I_{SD}(E,V_{GS},V_{DS}) dE~,
\end{align}
where, $V_{GS}$ and $V_{DS}$ are gate to source and drain to source voltages, respectively. $q$ is the elementry charge, $h$ is the Planck constant, $T(E,V_{GS},V_{DS}) = \sum_{k} T(E,k,V_{GS},V_{DS})$ is the transmission coefficient at Energy $E$ for a given bias ($V_{GS}$, $V_{DS}$), $\mu_{S/D}$ is the chemical potential at source/drain, and $f(E-\mu_{S/D})$ is the Fermi-Dirac distribution function at source/drain.\\
The $I_{SD}$ has three components, namely thermionic ($I_{thermal}$), source-to-drain tunneling ($I_{SDT}$) and BTBT ($I_{BTBT}$) currents. These components shown in Fig. \ref{fig:3} (b), can be expressed as, \\

\begin{subequations}
\begin{align}
I_{thermal}(V_{GS},V_{DS}) = \int_{-\infty}^{E_{V_{min},Ch}} I_{SD}(E,V_{GS},V_{DS}) dE~,\\
I_{SDT}(V_{GS},V_{DS}) = \int_{E_{V_{{min}},Ch}}^{E_{V,source}} I_{SD}(E,V_{GS},V_{DS}) dE~,\\
I_{BTBT}(V_{GS},V_{DS}) = \int_{E_{V,source}}^{\infty} I_{SD}(E,V_{GS},V_{DS}) dE~,
\end{align}
\end{subequations}
Here, $E_{V_{min},Ch}$ and $E_{V,source}$ are the minimum valance
band energy level in the channel and valance band energy level in the source. $I_{SD}(E,V_{GS},V_{DS})$ is current value at energy E for given bias $(V_{GS},V_{DS})$.

We use Poisson and NEGF modules implemented in NanoTCAD ViDES suites of codes\cite{NanoTCAD}. Periodic boundary conditions in the transverse direction are considered using 30 uniform wave vector points. Transport is assumed to be ballistic in nature due to short channel length used in this work. All transport simulations are performed at room temperature ($T = 300~K$).

\section{Results}
\subsection{Geometric and Electronic Structures}
\figref{fig:1} (a) shows the top view and (b) and (c) show side view along two different lateral directions of pentagonal PdSe$_2$ mono-layer. The mono-layer of pentagonal-PdSe$_2$ is buckled type and belongs to the space group $P2_1/c$. This mono-layer is a net of pentagonal networks, and each network consists of two Pd and three Se atoms (see Fig. \ref{fig:1} (a)). We start with the experimental bulk structure having lattice parameters a = 5.741 \AA, b = 5.886 \AA, and c = 7.691 \AA.
Then, we take out mono-layer structure from bulk and optimize it with 30~\AA~thick vacuum layer in the $\rm |z|$-direction. Comparison of optimized lattice constants (a and b) and bond lengths ($\rm d_{Pd-Se}$ and $\rm d_{Se-Se}$) with earlier works is shown in Table \ref{table1}. It's clear from Table \ref{table1} that our values are consistent with earlier results\cite{soulard,lan,deng,qin,oyedele}.

\figref{fig:1}~(d)~shows the electronic band structure of monolayer pentagonal-PdSe$_2$ from DFT (navy blue line) and TB (hollow green circle) along the high symmetry path ($\Gamma$-$\rm X$-$\rm S$-$\rm Y$-$\Gamma$). The red and black color circles in Fig. \ref{fig:1}~(d) indicate the location of conduction band minima (CBM) and valance band maxima (VBM), respectively. The band gap  value is 1.35 eV (indirect) and is very close to experimental value ($\rm E_{G,exp} \sim 1.3 \pm 0.2 eV$)\cite{oyedele}. The band structure from TB shows good agreement with DFT in the vicinity of VBM. The VBM lies on the way from $\Gamma$ to $\rm X$, hence the valley degeneracy is two. 
\figref{fig:1}~(e) shows the contour plot of the highest valance band. In the plot, symbols and black lines indicate the high symmetry points ($\Gamma$, $\rm X$, $\rm S$, $\rm Y$) and transport directions ([100] ($\Gamma$-$\rm X$) and [010] ($\Gamma$-$\rm Y$)), respectively.

\begin{figure}[!t]
	\centering
	\includegraphics[width=0.5\textwidth]{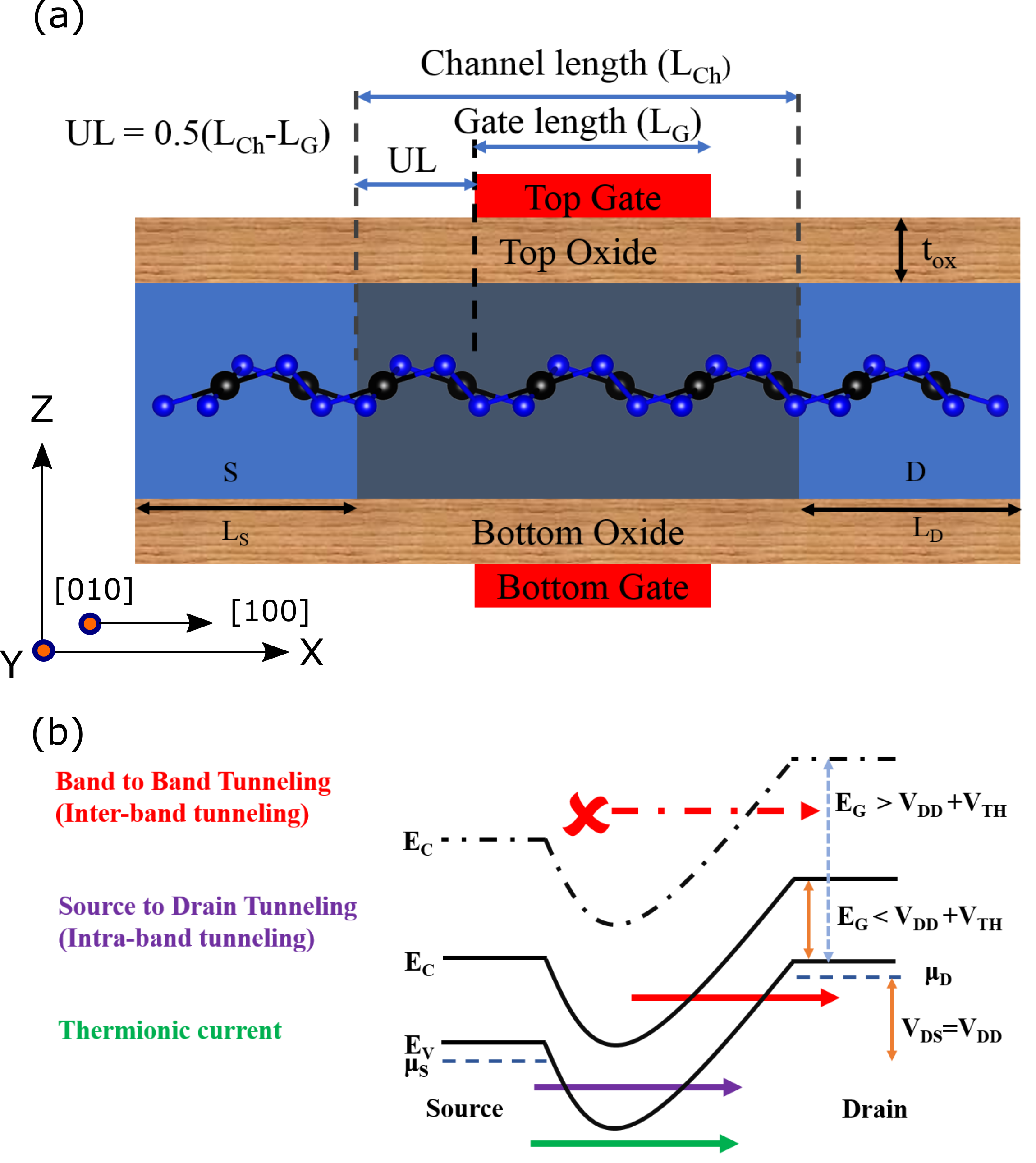}
	\caption{\textbf{Device Structure and Current Components:} (a) Schematic of the device under investigation. $\rm L_{Ch}$ and $\rm L_{S/D}$ are channel length and source/drain extension. $\rm L_G$ is gate length. UL is underlap length. $\rm t_{ox}$ is thickness of top and bottom gate oxide.~(b) The schematic of p-MOSFET band diagram in the OFF state for two different conditions of $\rm E_G$ ($\rm E_V$ is taken as reference): i) if $\rm E_G < V_{DD} + V_{T}$ (solid black line represents $\rm E_C$) and ii) if $\rm E_G > V_{DD} + V_{T}$ (dashed black line represent $\rm E_C$). The red, violet, and green arrows indicate the path for thermionic, SDT, and BTBT, respectively. A large $\rm E_G$ in this material ensures the condition $\rm E_G > V_{DD} + V_{T}$. Hence, BTBT is negligible in our devices.}
	
	\label{fig:3}
\end{figure}

\begin{table}[!b]
	\renewcommand{\arraystretch}{1.2}
	\caption{The optimized lattice constants (a and b) and bond lengths between Pd-Se and Se-Se for mono-layer pentagonal PdSe$_2$.  }
	\label{tab:example}
	\centering
	\begin{tabular}{|c|c|c|}
		\hline
		\rowcolor{lightgray} {Work} & a (\AA), b (\AA) &$\rm d_{Pd-Se}$ (\AA), $\rm d_{Se-Se}$ (\AA) \\
		\hline
		\hline
		
		This work & 5.748, 5.923 & (2.457, 2.469), 2.420  \\ 
		Soulard \it{et al.} \cite{soulard} (Exp.) &5.7457, 5.8679 &----\\
		
		Lan \it{et al.} \cite{lan} & 5.74, 5.91 &(2.45, 2.46), 2.41\\ 
		
		Deng \it{et al.} \cite{deng} &5.749, 5.495 &  ----\\ 
		
		Qin \it{et al.} \cite{qin} &5.7538, 5.9257 &  ----\\ 
		
		Oyedele \it{et al.} \cite{oyedele} &5.72, 5.93 &  ---- \\ 
		\hline
	\end{tabular}
	\label{table1}
\end{table}

\subsection{Electrical Characteristics and Performance Metrics}
\figref{fig:3}~(a) shows the schematic of the device. The device is a double gate (DG) MOSFET. The gate oxide is SiO$_2$, and the thicknesses of top and bottom gate oxides are the same. The channel material is pentagonal-PdSe$_2$. The source/drain is highly doped with p-type doping density of $2 \times 10^{17}~m^{-2}$. $\rm L_{Ch}$, $\rm L_G$, $\rm t_{ox}$, $\rm L_{S/D}$ and $\rm UL$, which are geometrical parameters of the device, represent channel length, gate length, oxide thickness, source/drain extension, and underlap length, respectively.

\begin{figure}[!t]
	\centering
	\includegraphics[width=0.5\textwidth]{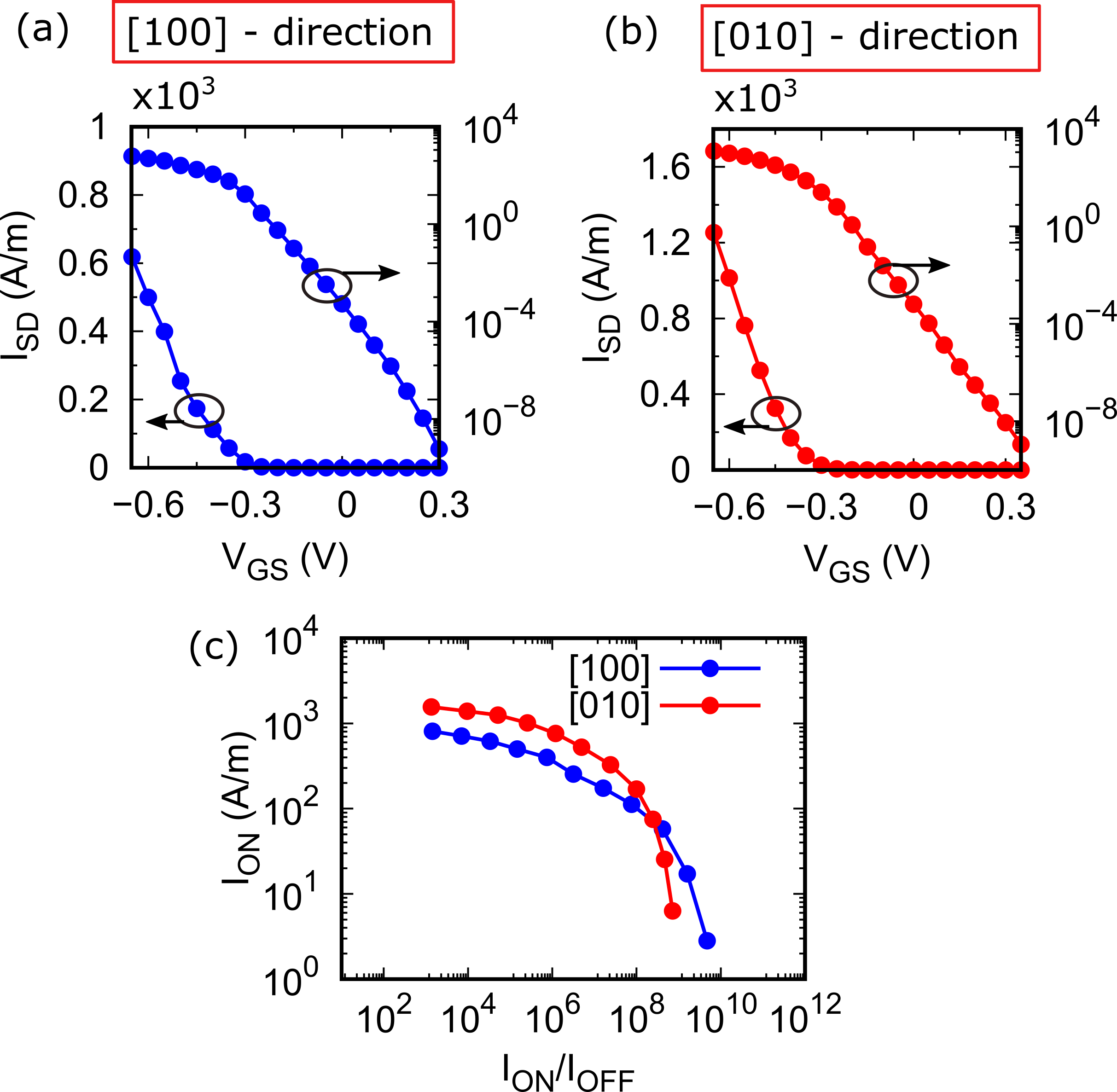}
	\caption{\textbf{Transfer Characteristics:} $I_{SD}$ vs $V_{GS}$ of FET for transport direction (a) [100] and (b) [010] with $\rm L_{Ch}$ $\sim$ 15.50 nm. The UL length and EOT are 0.0 nm and 0.6 nm, respectively. The power supply voltage ($\rm V_{DD}$) is 0.5V. (c) ON current ($\rm I_{ON}$) vs ON-OFF ratio ($\rm I_{ON}/I_{OFF}$). }
	
	\label{fig:4}
\end{figure}

\begin{table}[!b]
	\renewcommand{\arraystretch}{1.2}
	\caption{Performance comparison  of mono-layer pentagonal $\rm PdSe_2^{**}$ p-FETs  against International Roadmap for Devices and Systems (IRDS) $\rm 2020^{*}$\cite{ITRS2020} requirements for the year 2034.  }
	\label{tab:example}
	\centering
	\begin{tabular}{|c|c|c|c|c|c|c|}
		\hline
		\rowcolor{lightgray} & $\rm L_G$ &Transport &UL & 
		 &$\rm I_{ON}$  &$\rm I_{OFF}$ \\
		\rowcolor{lightgray} &(nm)&direction&(nm)& mV/decade& (A/m)&(A/m)\\
		
		\hline
		\hline
		& $\sim$ 15.50 & [100] & 0.00 &61 &618 &$10^{-1}$ \\
		& $\sim$ 15.50 & [100] & 0.00 &61 &463 &$10^{-2}$ \\
		& $\sim$ 15.50 & [100] & 0.00 &61 &176 &$10^{-4}$ \\
		
		\hline
		
		& $\sim$ 15.50 & [010] & 0.00 & 63& 1205&$10^{-1}$ \\
		& $\sim$ 15.50 & [010] & 0.00 & 63&883 &$10^{-2}$ \\
		& $\sim$ 15.50 & [010] & 0.00 & 63&348 &$10^{-4}$ \\
		
		\hline
		Y-2034  &  &  &  &  &  & \\
		HP & 12 & - & - & 70 & 1504 & $10^{-2}$\\
		LP & 12 & - & - & 65 & 861 & $10^{-4}$\\
		\hline
	\end{tabular}
	\label{table2}\\
	
	\footnotesize{*$\rm V_{DD}$ = 0.55 V; **$\rm V_{DD}$ = 0.50 V, EOT = 0.60 nm, and $\rm L_{S/D}$=10 nm.}\\
\end{table}

\hspace{0.01mm}
\subsubsection{Transfer Characteristics and the On Current}
\figref{fig:4} (a) and (b) show the transfer characteristics of the device for channel with transport directions along [100] and [010] axes. The channel length is 15.50 nm without underlap (i.e., $\rm L_{Ch} = L_{G}$), the supply voltage ($\rm V_{DD}$) is 0.50 V, and the source/drain extension region ($\rm L_{S/D}$) is 10 nm. For this channel length, source-to-drain tunneling (SDT) is negligible. The band-to-band tunneling (BTBT) is also negligible due to large band gap of this material. \figref{fig:4} (c) shows the ON current ($\rm I_{ON}$) vs. ON/OFF ratio ($\rm I_{ON}/I_{OFF}$) for both transport directions. $\rm I_{ON}$ is one of the important metrics for logic switches, because the maximum operating speed is proportional to $\rm I_{ON}$.
For $\rm I_{ON}/I_{OFF}$ $\sim$ $10^4$ ($10^5$), ON current for [010] direction $\rm I_{ON-[010]}$ ($\sim$ $1.38 \times 10^3 A/m$ ($1.14 \times 10^3 A/m$)) is approximately double than ON current for [100] direction $\rm I_{ON-[100]}$ ($\sim$ $6.86 \times 10^2 A/m$ ($5.23 \times 10^2 A/m$)). The reason for higher ON current in the [010] direction compared to the [100] direction is smaller effective mass  in the [010] direction, ($m_{[010]}^*\sim 0.25 m_o$) than [100] ($m_{[100]}^*\sim 0.40 m_o$).

We benchmark the performance of mono-layer pentagonal PdSe$_2$ p-FETs  against International Roadmap for Devices and Systems (IRDS) $\rm 2020$\cite{ITRS2020} requirements for the year 2034 in Table~\ref{table2}. According to the IRDS roadmap, option for the logic device is lateral gate-all-around (LGAA)-3D for the year 2034. The [010] ([100]) transport direction based p-FETs meets $\sim$ 41\% (21\%) and $\sim$ 60\% (31\%) requirement of ON current expected in IRDS roadmap for LP and HP applications, respectively.

\hspace{0.01mm}
\subsubsection{Channel Length Scaling}

\begin{figure}[!t]
	\centering
	\includegraphics[width=0.5\textwidth]{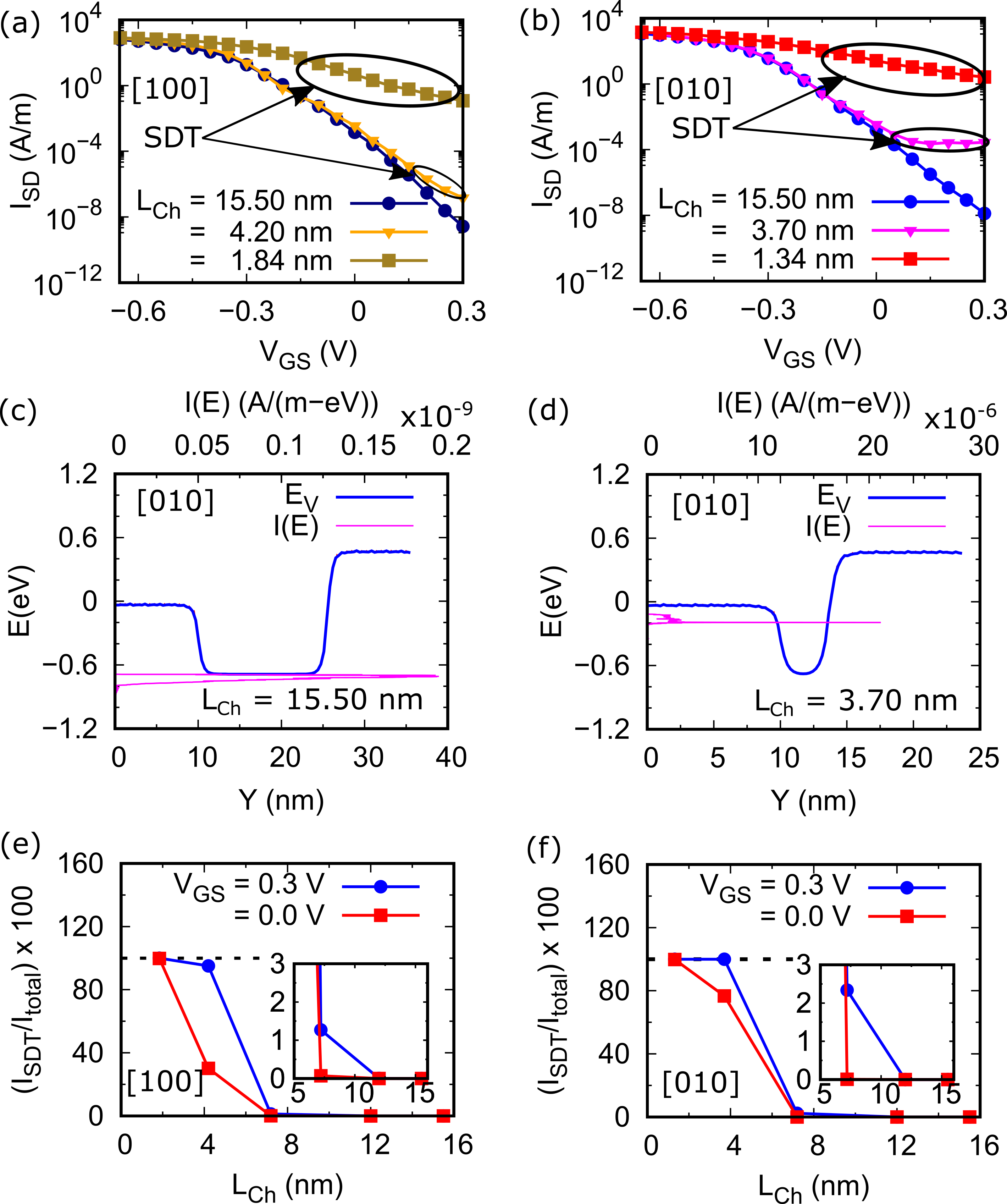}
	\caption{\textbf{Channel Length Scaling:} Transfer characteristics for different values of $\rm L_{Ch}$ (down to $\sim$ 1.5 nm) for transport directions (a) [100] and (b) [010]. Energy resolved current spectrum for (c) $\rm L_{Ch}$ = 15.50 nm and (d) $\rm L_{Ch}$ = 3.70 nm for transport direction [010]. The contribution percentage of $I_{SDT}$ to $I_{total}$ (= $I_{SDT} + I_{thermal}$) vs $\rm L_{Ch}$ for transport directions (e) [100] and (f) [010]. Insets in (e) and (f) show the contribution percentage for 5 nm $\rm <~ L_{Ch}~<$15.50 nm.}
	
	\label{fig:5}
\end{figure}

\begin{figure}[!t]
	\centering
	\includegraphics[width=0.5\textwidth]{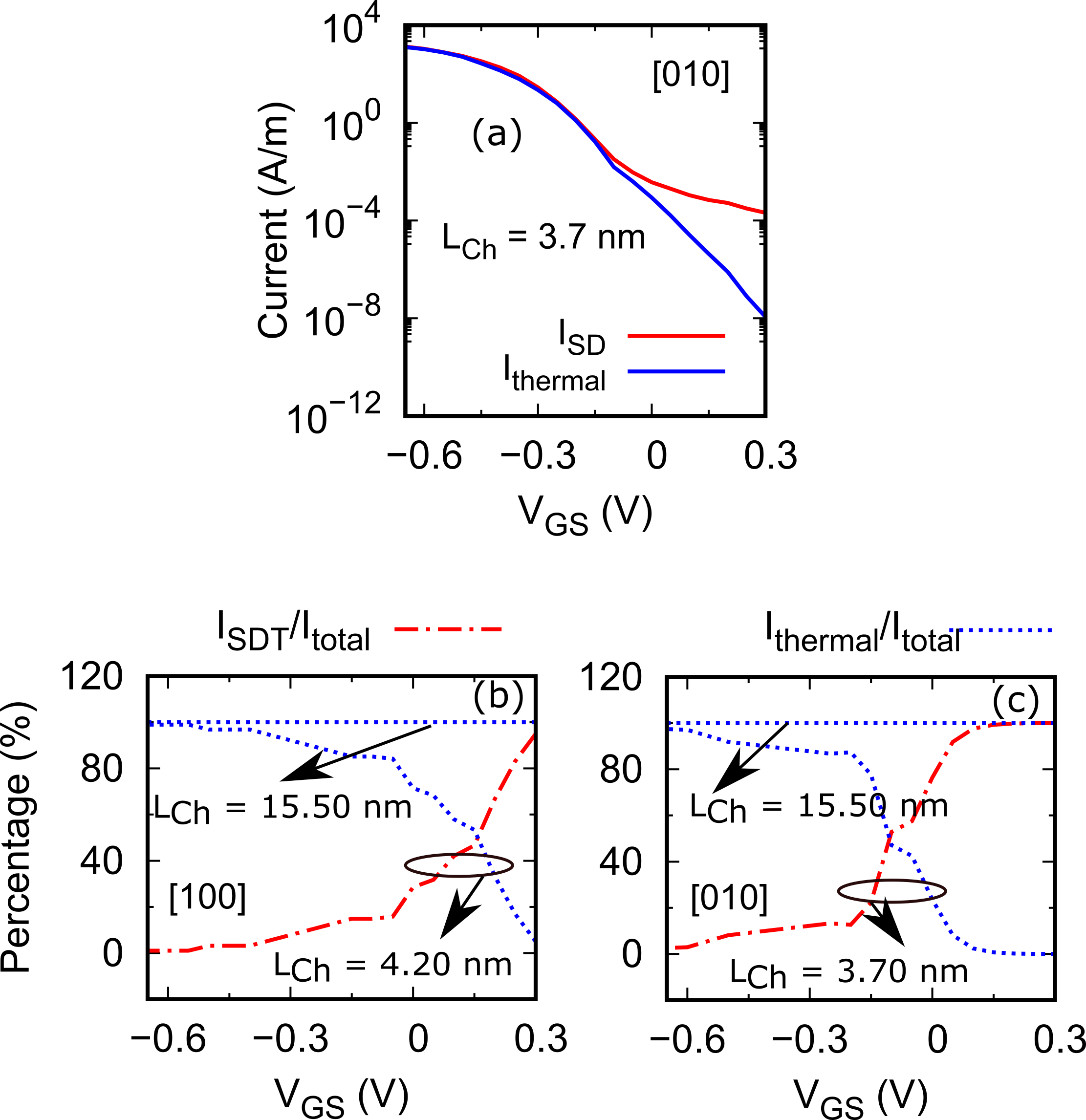}
	\caption{(a) $I_{thermal}$ and $I_{SDT}$ vs. $V_{GS}$, (b) and (c) contribution percentage of $I_{SDT}$ and $I_{thermal}$ to $ I_{total}$ vs. $V_{GS}$ for transport direction [100] and [010], respectively.} 
	
	\label{sdt_thermal}
\end{figure}

Now, we scale the channel length to study the impact of SDT on the device performance. The SDT decides the limit of gate length scaling\cite{sdt} and plays a significant role in determining the subthreshold slope (SS) and, hence OFF current ($\rm I_{OFF}$) at short channel length. \figref{fig:5} (a) and (b) show the transfer characteristics for different channel lengths with transport directions along [100] and [010]. The EOT, UL, and $\rm V_{DD}$ are 0.60 nm, 0.0 nm, and 0.50 V, respectively. The effect of SDT is more prominent for [010] transport direction than [100], because tunneling probability from source to drain is proportional to $exp{(- \sqrt{m^*})}$ \cite{griffiths} and $m_{[010]}^* < m_{[100]}^*$. \figref{fig:5} (c) and (d) show the energy resolved current spectrum and valance band for $\rm L_{Ch}$ = 15.50 and 3.70 nm, channel with transport direction [010] for ($V_{DS}$, $V_{GS}$) = (0.50, 0.30) V, which clearly indicate that $I_{SDT}$ majorly contributes to $I_{SD}$ for $\rm L_{Ch}$ = 3.70 nm. 
We plot the contribution percentage of $I_{SDT}$ to $I_{total}$ vs. $\rm L_{Ch}$ at $V_{GS}$ = 0.3 and 0.0 V in Fig. \ref{fig:5} (e) and \ref{fig:5} (f). For short channel lengths ($\sim$ 1.5 and 4.0 nm), the contribution of $I_{SDT}$ is $\sim$ 100\% to $I_{total}$ (i.e., $I_{total}$ $\simeq$ $I_{SDT}$) for both transport directions at $V_{GS}$ = 0.30 V. But, for $V_{GS}$ = 0.00 V and $\rm L_{Ch}$ $\sim$ 4 nm, it decreases to $\sim$ 30\% and $\sim$ 75\% for tansport directions [100] and [010], respectively. For $\rm L_{Ch}$ $>$ 10 nm, the contribution is $\sim$ 0\% (i.e., $I_{total}$ $\simeq$ $I_{thermal}$)  for both transport directions at $V_{GS}$ = 0.0 V and 0.3 V. For $V_{GS}$ = 0.3 V and $\rm L_{Ch}$ $\sim$ 7 nm, it is $\sim$ 1\% and 2.5\% for transport directions [100] and [010], respectively. Whereas, for $V_{GS}$ = 0.0 V and $\rm L_{Ch}$ $\sim$ 7 nm, the contribution is 0\% for both transport directions. \figref{sdt_thermal} (a) shows the thermionic part of total current and total current vs. $V_{GS}$ for transport direction [010] and $\rm L_{Ch}$ = 3.70 nm. The contribution percentage of $I_{SDT}$ to $I_{total}$ and $I_{thermal}$ to $I_{total}$ vs. $V_{GS}$ for transport directions [010] and [100] are shown in Fig. \ref{sdt_thermal} (b) and (c), respectively.  


The benchmarking of [010] transport direction (because this transport direction gives best performance) based p-FETs against ITRS 2013 requirements for the year 2028 is shown in Fig. \ref{table3}. For LP applications, the OFF state current is not meeting expected ITRS requirements. For HP applications, our device shows 15\% more ON current than expected in ITRS roadmap. Our simulations provide an upper-performance limit, as the transport is assumed to be ballistic in nature. However, the effect of phonon-scattering is very weak for short-channel devices. For $\sim$ 10 nm channel length, at least $\sim$ 80\% ballisticity has been shown for $\rm MoS_2$\cite{MoS2} and BP\cite{BP} based FETs. In fact, $> $ 90 \% ballisticity is reported for channel length below 5 nm\cite{MoS2}.

\begin{figure}[!b]
	\centering
	\includegraphics[width=0.5\textwidth]{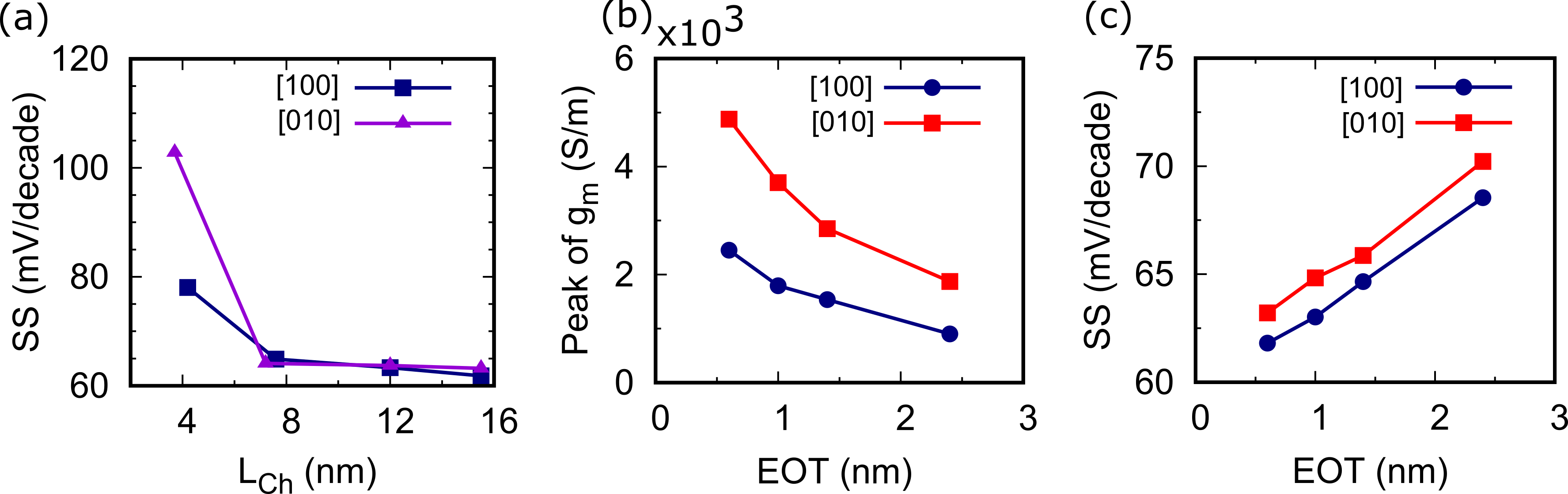}
	\caption{\textbf{Gate Control Parameters :} (a) SS vs $\rm L_{Ch}$.~(b) Peak transconductance vs EOT. (c)~SS vs EOT. }
	
	\label{fig:6}
\end{figure}

\hspace{0.01mm}
\subsubsection{Gate Control}
In addition to $\rm I_{ON}$, SS and $g_m$ are crucial figures of merit for device performance. They reveal the gate control in the sub-threshold and above-threshold region. Smaller value of SS indicates better gate control.

\begin{figure}[!t]
	\centering
	\includegraphics[width=0.5\textwidth]{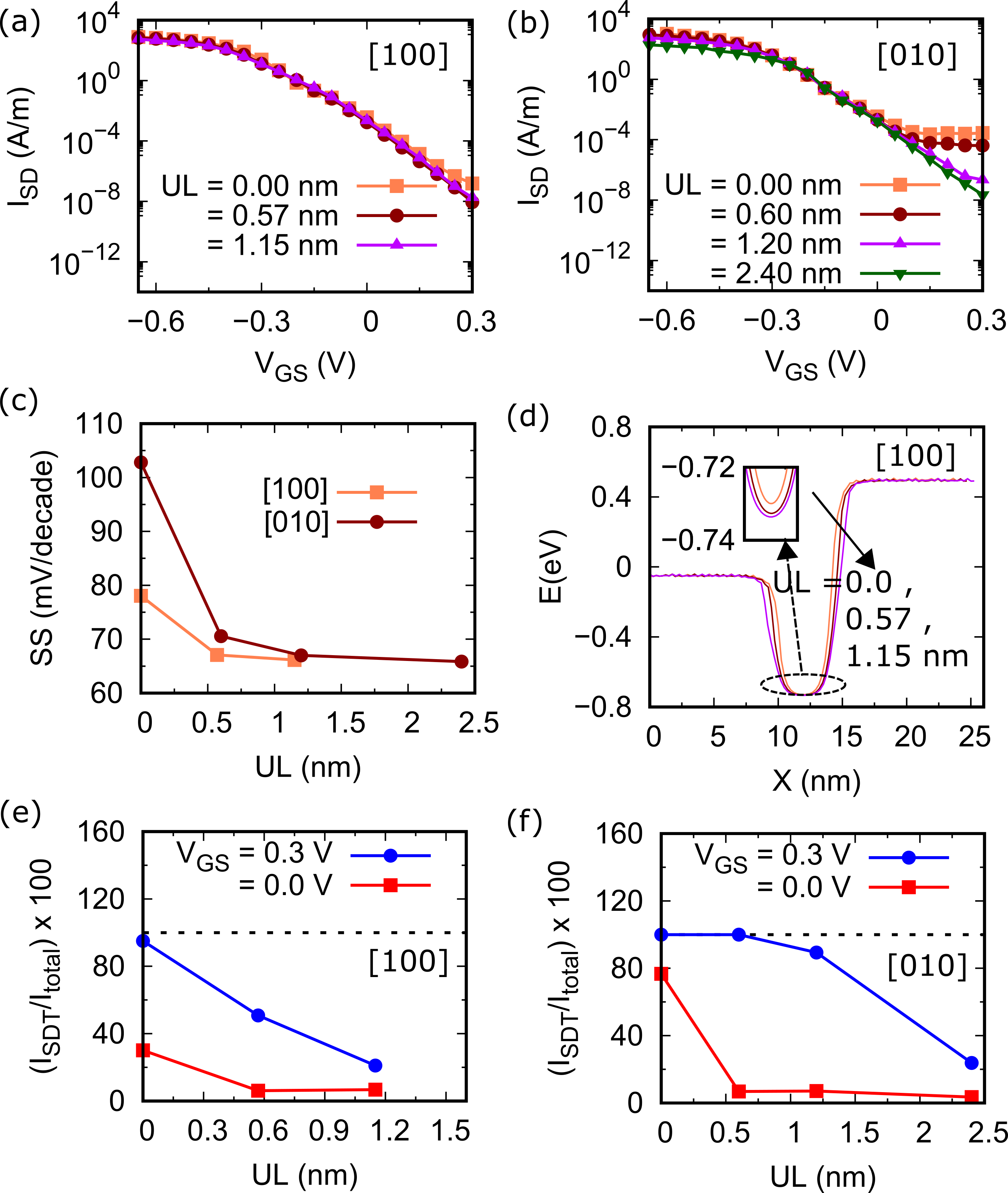}
	
	\caption{\textbf{Underlap Scaling :} Transfer characteristics for transport directions (a) [100] for $\rm L_G$ = 4.20 nm and (b) [010] for $\rm L_G$ = 3.70 nm for different UL lengths. (c) SS vs. UL length for the same $\rm L_{G}$ in (a) and (b). (d) Valance band diagram of device for transport direction [100] for different UL lengths. The contribution percentage of $I_{SDT}$ to $I_{total}$ vs UL lengths for transport directions (e) [100] and (f) [010]. EOT = 0.60 nm, $\rm V_{DD}$ = 0.50 V.}

	\label{fig:7}
\end{figure} 
\figref{fig:6} (a) shows that variation of SS with $\rm L_{Ch}$. The SS degradation in transport direction [010] is more prominent than [100]. The excellent switching characteristics (SS $<$ 70 mv/decade) are seen in both transport directions for $\rm L_{Ch}$ $>$ 7 nm. For $\rm L_{Ch} < 7~nm$, SDT plays a significant role in deciding OFF-current and SS (see Table \ref{table2}, Fig. \ref{fig:6} (a), \ref{fig:5} (e), and \ref{fig:5} (f)).

The variation of gate control parameters peak of $g_m$ and SS with EOT are shown in Fig. \ref{fig:6} (b) and (c). The $g_m$ is given by,
 
\begin{equation} \label{gm}
\begin{split}
g_m & = \frac{\partial I_{DS}}{\partial V_{GS}}, \\
\end{split}
\end{equation}
and, hence
\begin{equation} \label{gm_1}
\begin{split}
g_m& \propto v_h \frac{\partial Q_{Ch}}{\partial V_{GS}}.\\
\end{split}
\end{equation}
The density of states (D(E)) determines $\partial Q_{Ch}/\partial V_{GS}$ and is given by,
\begin{align} \label{DOS}
D(E) & = \frac{g_s g_v m_o}{2 \pi \hbar^2} \sqrt{m_{[100]}^* m_{[010]}^*}
\end{align}
where, $g_s$ and $g_v$ are spin and valley degeneracy. $Q_{Ch}$ is channel charge and $v_h$ is carrier velocity.\\
We find that both control parameters degrade almost linearly with EOT. The $g_m$ is higher for the [010] direction because small variation in $V_{GS}$ leads to same amount of change in channel charge (as D(E) is independent of channel orientation) for both transport directions and carrier velocity is more in [010] than [100] direction. In this work, we consider the gate as an ideal insulated electrode, but thin gate oxide leads to gate leakage current and hence, insulator thickness has a limit.

\hspace{-0.01mm}
\subsubsection{Underlap Scaling}
The tunneling width and the barrier height play vital roles in reducing the SDT at short channel length, which in turns leads to weaker gate control. Gate underlap\cite{UL_Ph1} structures overcome these shortcomings by increasing the effective channel length and consequently the tunneling width and the barrier height. But, a large underlap region can degrade the device performance, because the gate has better control underneath gate length region in channel compared to the UL region. Therefore, the UL length has to be optimized. \figref{fig:7} (a) and (b) show the transfer characteristics for different UL lengths for $\rm L_{G} \sim $ 4 nm for channel orientation along [100] and [010] directions. The SS vs UL length is plotted in Fig. \ref{fig:7} (c) to see the effect of UL length on gate control in sub-threshold region.

\begin{figure}[!t]
	\centering
	\includegraphics[scale=3]{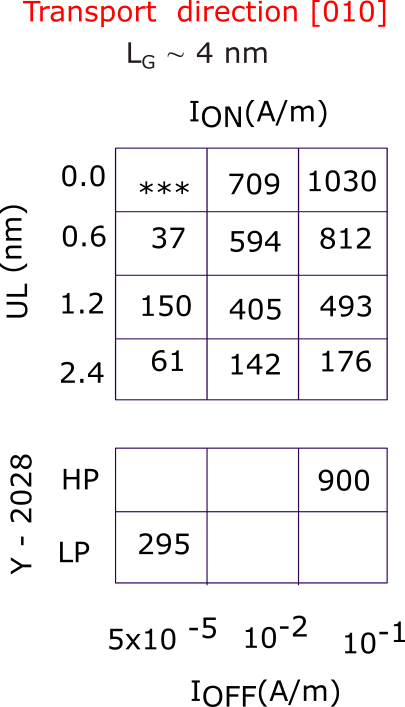}
	\caption{Performance comparison  of mono-layer pentagonal $\rm PdSe_2^{**}$ p-FETs  against $\rm ITRS~2013^*$\cite{ITRS2013} requirements for the year 2028. }

	\footnotesize{*$\rm V_{DD}$ = 0.64 V, EOT = 0.41 nm; **$\rm V_{DD}$ = 0.50 V, EOT = 0.60 nm, and $\rm L_{S/D}$=10 nm; ***$\rm I_{OFF}$ requirement can not be reached.}
	
	\label{table3}
\end{figure}

The impact of SDT in transport direction [100] is lower than in [010] and hence, in [100] channel, when UL length is increased from 0.0 to 1.15 nm, the contribution percentage of $I_{SDT}$ to $I_{total}$ reduces from $\sim$ 100\% to 30\% for $V_{GS}$ = 0.3 V and 30\% to 6\% for $V_{GS}$ = 0.0 V (see \figref{fig:7} (e)), which in turn reduces the SS from $\sim$ 78 to 66 mV/decade ($\rm I_{ON}$ drops by $\sim$ 30\%). However, in [010] channel, when UL is increased from 0.0 to 2.4 nm, the contribution percentage of $I_{SDT}$ to $I_{total}$ for $V_{GS}$ = 0.3 V reduces from $\sim$ 100\% to 30\%  and for $V_{GS}$ = 0.0 V, 30\% to 4\% (see \figref{fig:7} (f)), which in turn reduces the SS from $\sim$ 103 to 66 mV/decade ($\rm I_{ON}$ drops by $\sim$ 83\%).

\figref{table3} shows the benchmarking of p-FETs ($\rm L_G \sim 4nm$ and [010] transport direction) against ITRS 2013 requirements.
For LP applications: (i) without underlap, the OFF state current is not meeting expected ITRS requirements, (ii) introduction of underlap structure with UL = 0.60 nm and 1.20 nm meet $\sim$ 13\% and 51\% requirement of on current expected in ITRS roadmap, respectively, and (iii) further increasing the UL leads to decrease in on current. For HP applications, UL = 0.60 nm, 1.20 nm, and 2.40 nm meet $\sim$ 90\%, 55\%, and 20\% requirement of ON current expected in ITRS roadmap.

\section{conclusion}
We have studied the electronic structure and device characteristics of p-FETs based on mono-layers of pentagonal PdSe$_2$. We explicitly calculate the TB parameters of mono-layer PdSe$_2$ in the MLWF basis. Using these TB parameters we solve the coupled Poisson and Schr$\rm \ddot{o}$dinger equations, via recursive non-equilibrium Green's function formulation. We found that: (1) p-FETs show anisotropic transport behavior, (2) [010] oriented channel gives superior on-state performance than [100] oriented channel almost double ON current and $g_m$, (3) $\rm I_{ON}$ ($> 10^3$ A/m) is higher for transport direction [010] for $\rm I_{ON}/I_{OFF}$ $\sim$ $10^4$ and $10^5$, (4) Both transport directions show the sub-threshold slope less than 65 mV/decade, (5) the impact of SDT is more pronounced for the [010] oriented channel than [100] for very short channel length, (6) our device meets the HP ON current expected in ITRS 2013 roadmap for the year 2028, and (7) for LP, the upper performance limit of device meets $\sim$ 52\% ON current expected in ITRS 2013 roadmap for the year 2028. We expect that our findings may pave the way for realizing mono-layer pentagonal PdSe$_2$ based p-FETs in the near future.
\begin{table}[!t]
	\renewcommand{\arraystretch}{1.2}
	\caption{Comparison of p-FET with experimental data of Si based FinFET technology. $\rm I_{OFF}$ $\sim 10^{-1} A/m$.}
	\label{tab:example}
	\centering
	\begin{tabular}{|c|c|c|c|c|c|}
		\hline
		\rowcolor{lightgray} & $\rm L_G$ &$\rm V_{DD}$  & SS &$\rm I_{ON}$  &$\rm I_{ON}$ \\
		\rowcolor{lightgray} &(nm)&(V)& mV/dec.& (A/m)&/$\rm I_{OFF}$\\
		
		\hline
		\hline

		DG FinFET \cite{DGFinFET}&  10 & 1.2  & 125& 446  &4.46$\times 10^{3}$ \\
		GAA FinFET\cite{GAAFinFET}&  5 & 1.0 &208&497 &4.97 $\times 10^{3}$ \\
		PdSe$_2$ p-FET&  $\sim$ 4 & 0.5  & 103& 1030&$1.03 \times 10^{4}$ \\
		
		\hline

	\end{tabular}
	\label{table4}\\
\end{table}

\section{Acknowledgment}
K. Nandan thanks Prof. Amit Verma for fruitful discussions on electron transport, Barun Ghosh for help with first principles calculations, and Dr. Sarvesh S. Chauhan for his valuable feedback.


\bibliographystyle{IEEEtranDOI}
\bibliography{paper}

\end{document}